# Science Case and Requirements for the MOSAIC Concept for a Multi-Object Spectrograph for the European Extremely Large Telescope

C. J. Evans[1], M. Puech[2], B. Barbuy[3], P. Bonifacio[2], J.-G. Cuby[4], E. Guenther[5], F. Hammer[2], P. Jagourel[2], L. Kaper[6], S. L. Morris[7], J. Afonso[8], P. Amram[4], H. Aussel[9], A. Basden[7], N. Bastian[10], G. Battaglia[11], B. Biller[12], N. Bouché[13], E. Caffau[2], S. Charlot[14], Y. Clenet[15], F. Combes[16], C. Conselice[17], T. Contini[13], G. Dalton[18,19], B. Davies[10], K. Disseau[2], J. Dunlop[12], F. Fiore[20], H. Flores[2], T. Fusco[21], D. Gadotti[22], A. Gallazzi[23], E. Giallongo[20], T. Gonçalves[24], D. Gratadour[15], V. Hill[25], M. Huertas-Company[2], R. Ibata[26], S. Larsen[27], O. Le Fèvre[4], B. Lemasle[6], C. Maraston[28], S. Mei[2], Y. Mellier[14], G. Östlin[29], T. Paumard[15], R. Pello[13], L. Pentericci[20], P. Petitjean[14], M. Roth[30], D. Rouan[15], D. Schaerer[31], E. Telles[32], S. Trager[33], N. Welikala[18], S. Zibetti[23], B. Ziegler[34]

[1] UK Astronomy Technology Centre, Royal Observatory Edinburgh, Blackford Hill, Edinburgh, EH9 3HJ, UK
[2] GEPI, Observatoire de Paris, CNRS, Univ. Paris Diderot, Place Jules Janssen, 92190 Meudon, France
[3] Universidade de São Paulo, IAG, Rua do Matão 1226, Cidade Universitária, São Paulo, 05508-900
[4] Aix Marseille Université, CNRS, LAM UMR 7326, 13388, Marseille, France
[5] Thueringer Landessternwarte, Sternwarte 5, 07778 Tautenburg, Germany
[6] Astronomical Institute Anton Pannekoek, Amsterdam University, Science Park 904, 1098 XH, Amsterdam, The Netherlands
[7] Department of Physics, Durham University, South Road, Durham, DH1 3LE, UK
[8] CAAUL, Observatório Astronómico de Lisboa, Tapada da Ajuda, 1349-018 Lisbon, Portugal
[9] CEA, AIM, Service d'Astrophysique, Batiment 709, CE-Saclay, 91191 Gif sur Yvette, France
[10] Astrophysics Research Institute, Liverpool John Moores University, 146 Brownlow Hill, Liverpool L3 5RF, UK
[11] Instituto de Astrofísica de Canarias, calle Via Lactea s/n, 38200 San Cristobal de La Laguna, Tenerife, Spain
[12] Institute for Astronomy, Royal Observatory Edinburgh, Blackford Hill, Edinburgh, EH9 3HJ, UK
[13] Université de Toulouse, UPS-OMP, CNRS, IRAP, 9 Av. colonel Roche, BP 44346, 31028, Toulouse Cedex 4, France
[14] UPMC-CNRS, UMR7095, Institut d'Astrophysique de Paris, F-75014, Paris, France
[15] LESIA, Obs. de Paris, CNRS-UPMC, Univ. Paris-Diderot, PSL Research University, 5 place Jules Janssen, 92195 Meudon, France
[16] LERMA, Observatoire de Paris, CNRS, 61 Av de l'Observatoire, 75014 Paris, France
[17] School of Physics & Astronomy, University of Nottingham, University Park, Nottingham NG7 2RD, UK
[18] Astrophysics, Department of Physics, Keble Road, Oxford OX1 3RH
[19] Space Science and Technology, Rutherford Appleton Laboratory, HSIC, Didcot OX11 0QX
[20] INAF—Osservatorio Astronomico di Roma, via di Frascati 33, I-00040 Monteporzio, Italy
[21] ONERA, Optics Department, 29 avenue de la division Leclerc, 92320 Châtillon, France
[22] European Southern Observatory, Alonso de Córdova 3107, Vitacura, Casilla 19001, Santiago, Chile
[23] INAF-Osservatorio Astrofisico di Arcetri, Largo Enrico Fermi 5, 50125 Firenze, Italy
[24] Observatório do Valongo, UFRJ, Ladeira Pedro Antonio, 43, Saúde 20080-090 Rio de Janeiro, Brazil
[25] Observatoire de la Côte d'Azur, Laboratoire Lagrange, bd. de l'Observatoire, BP 4229, 06304 Nice Cedex 04, France
[26] Observatoire de Strasbourg, 11 rue de l'observatoire, 67000 Strasbourg, France
[27] Dept. of Astrophysics, IMAPP, Radboud University Nijmegen, PO Box 9010, 6500 GL Nijmegen, The Netherlands
[28] Institute of Cosmology and Gravitation, University of Portsmouth, Burnaby Road, Portsmouth, PO1 3FX, UK
[29] Dept. of Astronomy, Oskar Klein Centre, Stockholm University, AlbaNova University Centre, SE-106 91 Stockholm, Sweden
[30] Leibniz-Institut für Astrophysik Potsdam, An der Sternwarte 16, 14482 Potsdam, Germany
[31] Geneva Observatory, Université de Genève, 51 chemin des Maillettes, 1290, Versoix, Switzerland
[32] Observatorio Nacional, R. Gal. Jose Cristino 77, São Cristóvão, Rio de Janeiro, RJ 20921-400, Brazil
[33] Kapteyn Astronomical Institute, University of Groningen, Postbus 800, 9700 AV, Groningen, the Netherlands
[34] Dept. of Astrophysics, University of Vienna, Türkenschanzstrasse 17, 1180 Wien, Austria

## ABSTRACT

Over the past 18 months we have revisited the science requirements for a multi-object spectrograph (MOS) for the European Extremely Large Telescope (E-ELT). These efforts span the full range of E-ELT science and include input from a broad cross-section of astronomers across the ESO partner countries. In this contribution we summarise the key cases relating to studies of high-redshift galaxies, galaxy evolution, and stellar populations, with a more expansive presentation of a new case relating to detection of exoplanets in stellar clusters. A general requirement is the need for two observational modes to best exploit the large ($\geq$40 arcmin$^2$) patrol field of the E-ELT. The first mode ('high multiplex') requires integrated-light (or coarsely resolved) optical/near-IR spectroscopy of >100 objects simultaneously. The second ('high definition'), enabled by wide-field adaptive optics, requires spatially-resolved, near-IR of >10 objects/sub-fields. Within the context of the conceptual study for an ELT-MOS called MOSAIC, we summarise the top-level requirements from each case and introduce the next steps in the design process.

## 1. INTRODUCTION

The workhorse instruments of the 8-10m class observatories have become their multi-object spectrographs (MOS), providing comprehensive follow-up to both ground-based and space-borne imaging. With the growing amount of deep imaging surveys from, e.g., the *Hubble Space Telescope (HST)* and the VISTA telescope, there are a plethora of spectroscopic targets which are already beyond the sensitivity limits of current facilities. This wealth of targets will grow even more rapidly in the coming years, e.g., full operations of ALMA, the launches of the *James Webb Space Telescope (JWST)* and *Euclid* missions, and the advent of the Large Synoptic Survey Telescope (LSST) and the Square Kilometre Array (SKA). Thus, a key requirement underlying plans for the next generation of ground-based telescopes, the Extremely Large Telescopes (ELTs), is for even greater sensitivity for optical and infrared (IR) spectroscopy. Moreover, with only three ELTs planned, and given their large construction and operational costs, the need to make efficient use of their focal planes will become even more compelling than at current facilities.

A first summary of the ELT-MOS case and the related top-level instrument requirements was presented by Evans et al. (2012), with more expansive discussion given in the '*ELT-MOS White Paper*' by Evans et al. (2013). Over the past 18 months we have revisited the MOS science case via updates to the existing cases and inclusion of new topics. These efforts will be presented at length in v2 of the White Paper (to be released later in 2014); in this contribution we summarise the main updates/additions compared to the overview from Evans et al. (2012) and v1 of the White Paper.

MOS observations will be strongly influenced by the adaptive optics (AO) capabilities of the E-ELT, which has been designed with a large adaptive mirror (M4) and a fast tip-tilt mirror (M5) to correct for turbulence in the lower layers of the atmosphere via ground-layer adaptive optics (GLAO); higher-performance AO will be delivered by dedicated modules or in the instruments. Evans et al. (2012) therefore identified two observational modes for ELT-MOS targets:

- *High-definition mode:* Observations of tens of channels at fine spatial resolution, with multi-object adaptive optics (MOAO) providing high-performance AO for selected sub-fields (e.g. Rousset et al. 2010);
- *High multiplex mode:* Integrated-light (or coarsely resolved, via GLAO) observations of >100 objects.

The instrument requirements which flow down from these cases/modes are now being used to inform the conceptual design of the Multi-Object Spectrograph for Astrophysics, IGM, and Cosmology (MOSAIC; see Hammer et al. these proceedings). The MOSAIC consortium combines the expertise and experience of groups that worked on past Phase A studies of MOS concepts for the E-ELT (namely: EAGLE, Cuby et al. 2010; OPTIMOS-DIORAMAS, Le Fèvre et al. 2010; OPTIMOS-EVE, Navarro et al. 2010), together with new partners. We note that the E-ELT has been designed to provide a 10 arcmin diameter focal plane but that the infrastructure needed for guide stars (AO/tracking) will obstruct the outer parts of the field. In the following we adopt the same focal-plane area (40 arcmin$^2$, equivalent to a 7 arcmin diameter field) as for the EAGLE Phase A design, although we will revisit this in the course of the MOSAIC study.

By way of background context, ESO have selected two first-light instruments for the telescope: MICADO, a near-IR imager (and potentially a single-object spectrograph, R. Davies et al. 2010) and HARMONI, a (red-)optical/near-IR integral-field spectrograph (Thatte et al. 2010). Both instruments will exploit the best image quality of the E-ELT via high-performance AO but will be limited in their spatial extent on-the-sky; i.e., a ~0.8 arcmin$^2$ field-of-view imager, and a monolithic (i.e. single-target) integral-field unit (IFU). HARMONI will be well suited to spectroscopy of individual high-$z$ galaxies of interest, and stars in very dense regions (e.g. inner parts of spirals, cluster complexes). However, the *large* samples needed to explore galaxy evolution, both at high and low redshifts, will require MOS observations.

## 2. UPDATES TO THE ELT-MOS SCIENCE CASE

We have identified eight key science cases (SC) for an ELT-MOS, which form the core cases for MOSAIC:

- SC1: 'First light' – Spectroscopy of the most distant galaxies;
- SC2: Evolution of large-scale structures;
- SC3: Mass assembly of galaxies through cosmic time;
- SC4: Active Galactic Nuclei (AGN)/Galaxy co-evolution & AGN feedback;
- SC5: Resolved stellar populations beyond the Local Group;
- SC6: Galaxy archaeology;
- SC7: Galactic Centre science;
- SC8: Planet formation in different environments.

The broad theme of the first seven is charting galaxy evolution over cosmic time, from resolved stellar populations in nearby galaxies out to observations of the most distant galaxies. In addition, SC8 draws on the latest results in the quickly growing field of exoplanet research. We now go through each of the cases, briefly highlighting updates since Evans et al. (2012, 2013) and introducing the new cases (SC4, 7, and 8).

**2.1. SC1: 'First Light' – Spectroscopy of the most distant galaxies**

For a complete picture of the star-formation and quasar activity responsible for reionisation of the early Universe we need an inventory of the first galaxies. In short, analysis of Ly-α emitters and Lyman-break galaxies (LBGs) at different redshifts, via their emission lines and luminosity functions. This will constrain when reionisation occurred and identify the galaxies responsible. However, spectroscopic confirmation of photometrically-selected targets at the highest redshifts is rare, and becomes essentially impossible at $z \geq 7$ with current facilities, particularly if the observed decrease in Ly-α emission at the largest redshifts is confirmed.

Deep observations of faint continuum-selected sources with MOSAIC will provide the necessary observational constraints to robustly determine the Ly-α properties and hence the ionisation state of the inter-galactic medium (IGM), from $z$=5 to 13. Such a programme will also determine the properties of these first, bright galaxies including their mass function and details of their interstellar media (ISM), outflows, and stellar populations, via analysis of their continuum/absorption-line, rest-frame UV spectra (e.g., Fig. 1). This will complement the high-quality spectral energy distributions expected from *JWST* imaging (which will be more limited in its spectroscopic capabilities, particularly for continuum observations, compared to the E-ELT). Simulation work is now underway to determine the optimum spatial pixel size, in terms of delivered S/N, for such observations with MOSAIC (e.g. Fig. 1; Disseau et al. these proceedings).

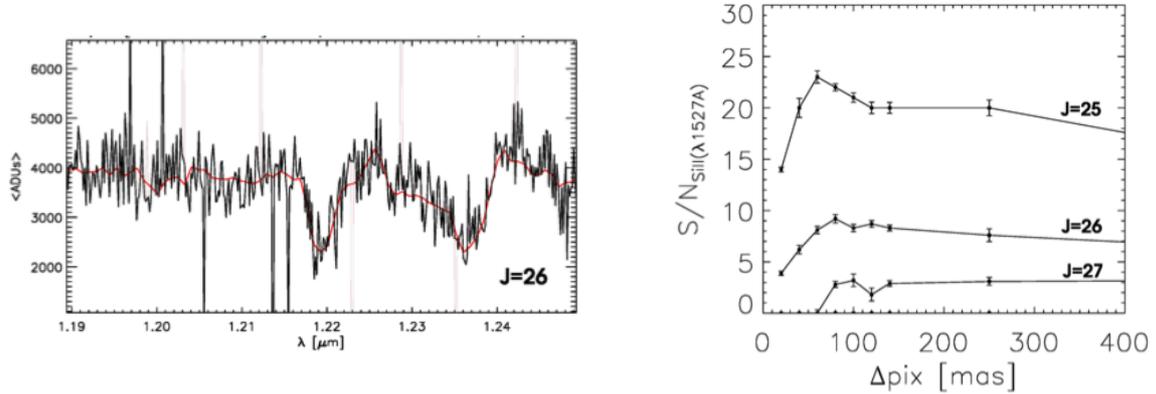

**Figure 1:** *Left:* Simulated spectrum of rest-frame SiII λ1527Å absorption in a $z$ = 7 galaxy (S/N > 5, total integration of 40 hrs, spatial pixel (on-the-sky) of 80 mas, $J_{AB}$ = 26 mag). *Right:* Summary of simulated performances as a function of spatial pixel size and *J*-band magnitude – the optimal spatial pixel size for this example simulation is 60-100 mas.

*Finding and following-up the targets:* The latest *HST* observations are already revealing sources which are beyond our spectroscopic capabilities, and we can expect additional high-$z$ candidates coming from the *JWST* and *Euclid* missions and from MICADO imaging. Other new projects will probably also emerge in the coming years such as, e.g., the *Wide-field Imaging Surveyor for High-redshift (WISH)* mission concept now under study for the Japanese Aerospace Exploration Agency (JAXA), which aims to reach ~1.5 mag deeper in the near-IR than the *Euclid* Deep Survey. Illustrative source densities expected for high-$z$ targets from *Euclid* and *WISH* are summarised in Table 1.

**Table 1:** Expected sources per 40 arcmin$^2$ field, as a function of redshift, for the *Euclid* Deep and potential *WISH* surveys.

|  |  | $5 < z \leq 6$ | $6 < z \leq 7$ | $7 < z \leq 8$ | $8 < z \leq 9$ |
|---|---|---|---|---|---|
| *Euclid* **Deep** | $J_{AB} \leq 26$ | 13.68 | 3.14 | 0.45 | 0.16 |
| *Euclid* **Deep** | $H_{AB} \leq 26$ | 13.68 | 3.14 | 0.45 | 0.21 |
| *WISH* **Deep** | $H_{AB} \leq 27.4$ | 127 | 60 | 21 | 12 |

Notes: Quoted depths are 5σ point-source sensitivities; the expected densities from *Euclid* in the *J*- and *H*-bands are identical except for the last redshift bin (as the Lyman break enters the *J*-band at $z$=8). These numbers were derived using the UV luminosity functions from Bouwens et al. (2007; at z=3.8, 5, 5.9) and McLure et al. (2013; at z=7, 8, 9).

## 2.2. SC2: Evolution of large-scale structures

This case has been expanded beyond those given by Evans et al. (2012, 2013) and now includes three topics:

*Tomography of the IGM:* This was the large-scale structure case discussed by Evans et al. (2012, 2013) and involves mapping gas in the IGM via observations of the redshifted Ly-α forest along sightlines to distant quasars or LBGs. This case leads to a third observational mode, for GLAO-corrected/seeing-limited, blue optical IFU spectroscopy (extending shortwards to at least 0.4 μm) of ≥10 sightlines across the MOSAIC patrol field.

*High-redshift clusters:* Rich samples of clusters and proto-clusters of galaxies have been discovered in recent years via selection as red or star-forming over-densities (e.g. Mei et al. 2014, and references therein). Of particular interest are systems at $z > 1.5$, as this is the critical epoch for the assembly/formation of their galaxy populations, and which will evolve into the local cluster populations we see today. Studies of the properties of these systems is therefore important for our understanding of cluster assembly, but also for its impact on the formation and evolution of early-type galaxies. Such high-$z$ systems generally appear to be undergoing star formation, but some clusters at $z \sim 2$ already have a well-defined red sequence, with a predominance of passive galaxies (Zeimann et al. 2012; Gobat et al. 2013; Newman et al. 2014). This varied behaviour suggests that the early-type populations in the most massive clusters were already quenched at $z > 2$, and that their progenitors were star-forming galaxies at $z \sim 2.5$-5, in which we start to detect proto-clusters from far-IR observations and emission-line surveys (Capak et al. 2011; Tadaki et al. 2014; Tan et al. 2014).

We require ELT spectroscopy to follow-up these early-type progenitors in primordial halos (up to $z \sim 5$), in which we need to observe the full spatial extent of the halos that will evolve into present-day clusters in the local Universe (in which each halo will host a few up to tens of galaxies). Current scenarios of early-type galaxy formation propose both galaxy mergers and disk accretion. To understand the ratios of early-type progenitors formed by these different modes, we will study their internal dynamics using redshifted emission lines (Hα, [OII], [OIII]); the ratios of these will also enable us to separate out AGN and to study the stellar populations in the galaxies (linking to SC3 and SC4).

*Galaxy cluster dynamics:* Moving closer to home, massive galaxy clusters are found to contain an important diffuse luminous component: 'intracluster light' (ICL). This is comprised of stars external to galaxy halos in the intracluster environment (e.g. Arnaboldi et al. 2002, Burke et al. 2012, Guennou et al. 2012) and can be used to study the dynamical history of the clusters as these stars have been tidally stripped by galaxy interactions (e.g. Rudick et al. 2011).

The ICL is only revealed with deep imaging, and groups have recently started to investigate its distribution in clusters as a function of redshift, e.g., CL0024+17 at $z = 0.4$ (Fig. 2; Giallongo et al. 2014). Model predictions in CDM scenarios show that the ratio of the total stellar mass lost by stripping compared to that which remains in the galaxies depends most strongly on cluster properties such as the total mass and dark matter distribution; in the example of CL0024+17, the predicted ICL fraction as a function of radius is in broad agreement with the observations (see Giallongo et al. 2014). If the ICL is produced via tidal stripping we would also expect a less significant component (and fewer faint dwarf galaxies) in the cluster core compared to the outer regions; hints of this effect are seen in the example of CL0024+17. Although deep imaging is very powerful for these studies, we currently lack spectroscopy, of both the ICL component and the dwarf galaxies in the clusters, to determine the dynamical properties of the clusters. In this respect, an ELT-MOS will be essential to probe the tidal and merging dynamics in galaxy cluster cores up to $z \sim 1.5$.

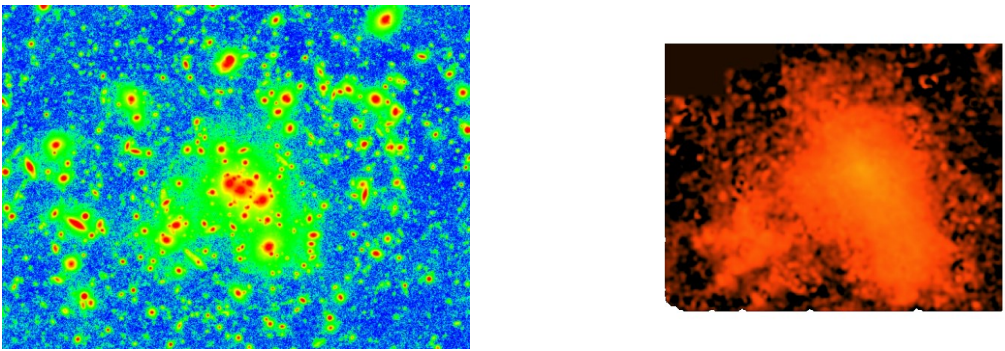

**Figure 2:** *Left:* Central 3×2.4 arcmin of the CL0024+17 cluster ($z = 0.4$). *Right:* Intracluster light (ICL) in CL0024+17 after removal of the galaxy contributions (see Giallongo et al. 2014).

## 2.3. SC3: Mass Assembly of Galaxies Through Cosmic Time

Little is known about the early evolution of massive galaxies and, in particular, regarding the relative impact of in-situ (star formation) vs. external events (mergers, large-scale environment) in their subsequent growth of mass and size. In the White Paper we have focussed on three examples of galaxy evolution cases which require ELT-MOS spectroscopy:

*Spatially-resolved spectroscopy of high-z emission-line galaxies:* The use of IFUs on 8-10m class telescopes has heralded a new era of galaxy studies, in which the spatially-resolved kinematics and physical properties can be determined for massive emission-line galaxies from z ~0.4 to 3 (e.g., see review by Glazebrook, x2013). In particular, the use on AO with instruments such as VLT-SINFONI and Keck-OSIRIS has provided unprecedented views of the substructure and properties of high-z galaxies (e.g. Fig. 3). However, at redshifts of $z > $ ~1.5, only the most massive, high-surface-brightness systems are within the grasp of current facilities, and most of these samples have been assembled (necessarily) from a range of selection criteria. Whether these samples are representative of the early Universe remains highly uncertain, and there are a wide range of opinions in the community as to which processes drive galaxy evolution. Only with a large observational sample of *mass selected* emission-line galaxies, mapping their spatially-resolved physical and chemical properties, will we make significant progress in this field. (In addition to the discussion of this case in the White Paper, we also direct the reader to the simulations by Puech et al. 2008, 2010.)

*Stellar populations in high-z galaxies:* Insights into the star-formation and assembly histories of galaxies at $z > 1$ can also be obtained from absorption-line spectroscopy. Star-formation rates and ages can be estimated via their redshifted 4000Å breaks ($D_{4000}$) and higher-order Balmer lines (H$\delta$, H$\gamma$ and H$\beta$), while metallic features (from iron and a variety of $\alpha$-elements) can be used to investigate chemical enrichment and its typical timescales (via the [$\alpha$/Fe] ratio). Massive, quiescent galaxies at $z$ ~2 have typical magnitudes of $m_{AB} >21$ so, even with long exposures, current telescopes can only provide marginal S/N for even the brightest galaxies at $z$ ~2 (e.g. Kriek et al. 2009; Toft et al. 2012). To observe large samples of the fainter and less massive galaxies we require MOS observations with an ELT; these observations only require integrated-light observations so seeing-limited/GLAO observations are sufficient.

*The puzzling role of high-z dwarfs in galaxy evolution:* If hierarchical models of galaxy assembly are correct we expect the fraction of galaxies which are dwarfs to increase as a function of redshift, manifested as a steepening of the galaxy luminosity/mass function. Such an increase is seen from rest-frame UV data (e.g. Ryan et al. 2007; Reddy & Steidel, 2009), and in the stellar mass function (Santini et al. 2012; Duncan et al. submitted; Grazian et al. submitted), but these are largely determined from photometric redshifts. At high redshifts the faint-end slope might be expected to be influenced by denser environments (e.g., Khochfar et al. 2007), but at lower redshifts such a trend is not seen. A key test is to obtain spectra of the plethora of intrinsically faint galaxies at $z > 1$ discovered with the *HST* (Ryan et al. 2007). Given the faint magnitudes of these galaxies ($m_{AB}$ ~24-27) only MOSAIC will be able to characterise this galaxy population, which might be an important but unexplored reservoir of baryons at high redshift.

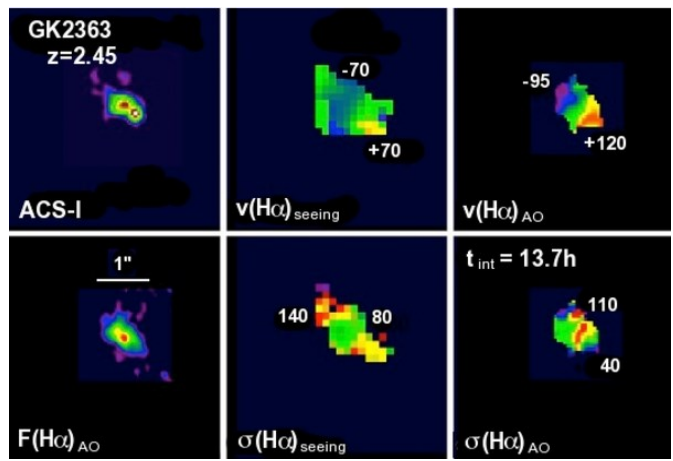

**Figure 3:** An example of state-of-the-art of IFU spectroscopy at $z$ = 2.45 from VLT-SINFONI observations (Newman et al. 2013). The upper-left panel shows an *I*-band *HST* image, with H$\alpha$ velocity maps from seeing-limited and AO-corrected observations shown in the remaining upper panels; much of the velocity information is lost without AO observations.

## 2.4. SC4: AGN/Galaxy Co-evolution and AGN Feedback

One of the most important puzzles in cosmology is the combined evolution and growth of galaxies and their central super-massive black holes (SMBHs). A key aspect is how galaxy and SMBH growth is self-regulated by feedback processes which are thought to arise from AGN and SNe-driven outflows. These could heat up the ISM and expel it from the host galaxy, therefore playing a major role in galaxy evolution (e.g. Croton et al. 2006; Menci et al. 2008; Zubovas & King 2012). For example, such outflows are thought to blow away most of the reservoir of cold gas, thus terminating galaxy growth and creating a population of 'red-and-dead', gas-poor galaxies.

Evidence for such AGN-driven outflows has been growing rapidly in recent years (e.g., Villar-Martin et al. 2011; Oh et al. 2013), such that they are thought to be a common feature that will affect the host galaxies of most AGN. The most important characteristics to determine for these outflows are their physical sizes and their mass-outflow rates, but with current facilities we have only been able to measure these for approx. two dozen of the brightest sources (i.e. nearby AGN or extremely luminous high-$z$ AGN), e.g. Fig. 4 (from Cano-Diaz et al. 2012).

The bulk of the AGN population remains out-of-reach at present. In particular, normal L* AGN at $z = 1$-$3$, i.e., the peak of both AGN and galaxy evolution, are effectively impossible to study with current facilities. The E-ELT will provide spectroscopy of faint AGN with bolometric luminosities of $\sim 10^{44}$-$10^{45}$ ergs/s, which is close to (or even below) L* at these redshifts. Moreover, ELT spectroscopy will provide additional diagnostics such as broad [Ne V] emission, Al III absorption, and other UV absorption features. This would enable the search and characterisation of ionised-gas outflows at significantly larger redshifts than at present, as illustrated by Fig. 4. Specific questions that will be addressed by MOSAIC spectroscopy include:

- What are the morphologies, masses, mass-flow rates and kinetic energies of the observed outflows?
- What are the best tracers to obtain reliable estimates of the outflow properties?
- What is the relation between AGN- and SNe-driven winds? Which is dominant?
- How common are massive outflows at the peak epoch of AGN and galaxy assembly ($z = 1$-$3$)?
- In which galaxy/AGN types do massive outflows occur? And in which phase of the evolution of a galaxy?
- How long does the active feedback phase last?
- How are molecular outflows linked to ionized outflows?
- How do outflow properties (masses, mass-loss rates, sizes) correlate with other AGN/galaxy properties?

Note that the full capabilities of ALMA will expand the discovery space for molecular outflows hugely in the future; the E-ELT will provide strong synergies with ALMA, leading to similar gains for studies of ionized/neutral gas outflows.

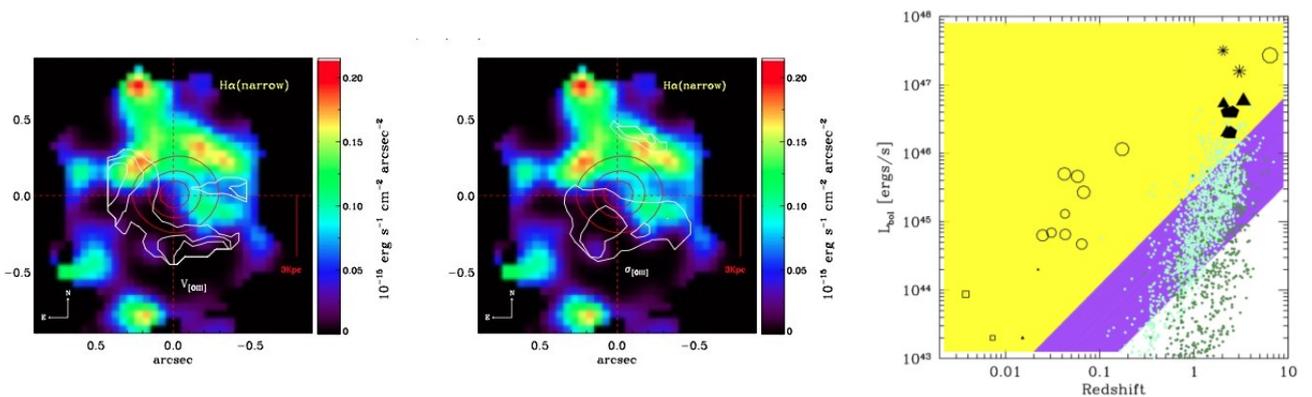

**Figure 4:** *Left/Centre:* VLT-SINFONI integral-field spectroscopy of a luminous quasar at $z = 2.4$; star formation (as traced by Hα) appears suppressed in the SE region where the strongest outflow is traced by [OIII] emission (Cano-Diaz et al. 2012). *Right:* $L_{bol}$–redshift plane for AGN with published outflow properties (black points). AGN from the *Chandra*-COSMOS and CDFS surveys are indicated by the cyan/green points. The yellow area is the region of parameter space accessible with current facilities; the E-ELT will enable observations in the purple region, providing a huge gain in studies of AGN outflows.

## 2.5. SC5: Resolved Stellar Populations Beyond the Local Group

With its vast primary aperture and excellent angular resolution, the E-ELT will be *the* facility to unlock spectroscopy of evolved stellar populations in the broad range of galaxies in the Local Volume, from the edge of the Local Group, out to Mpc distances. This will bring a wealth of new and exciting target galaxies within our grasp – spanning a broader range of galaxy morphologies, star-formation histories and metallicities than those available to us at present – enabling us to test theoretical models of galaxy assembly and evolution. There are many compelling target galaxies for stellar spectroscopy of individual resolved stars with the E-ELT, including the spiral dominated Sculptor 'Group' (at 2-4 Mpc) and the M83/NGC5128 (Cen A) grouping (at ~4-5 Mpc). There is already substantial deep imaging available of these galaxies from the *HST* and ground-based telescopes, i.e. we already have catalogues of potential targets, but lack the spectroscopic sensitivity with 8-10m telescopes for sufficient S/N. HARMONI will be well suited to spectroscopy of stars in individual regions in external galaxies (and the Milky Way), but the larger samples needed to explore entire galaxy populations will require observations with MOSAIC.

The E-ELT will also be important for spectroscopy of stellar populations in more distant galaxies. In terms of spatial resolution, observations of their evolved stellar populations will move from 'resolved' to 'semi-resolved' (building a bridge toward the integrated-light studies from SC3). Studies in targets such as NGC 1291 (8 Mpc), NGC 4594 (the Sombrero, ~10 Mpc), NGC 3379 (~11 Mpc) and systems in the Virgo Cluster will shed light on important questions relating to large-scale structure (e.g. Gadotti & Sanchez-Janssen, 2012). Additional motivation to reach targets at >10 Mpc is provided by the need for detailed results in systems at the higher end of the galaxy mass distribution (e.g. M87), enabling us to investigate the formation of bulges in disk galaxies over the full mass spectrum.

For determinations of stellar metallicities and radial velocities (RVs), we envisage two different diagnostic approaches. If stellar crowding is not a limiting factor, GLAO-corrected observations will suffice and the primary diagnostic will likely be the calcium triplet (at 0.85-0.87μm); the correlation of its intensity with metallicity is well understood (e.g. Cole et al. 2004; Carrera et al. 2007; Battaglia et al. 2008) and appears robust to metallicities as low as [Fe/H] ≥ −4 (Starkenburg et al. 2010). In denser regions of external galaxies we will require the improved spatial resolution obtained from high-performance AO. In these cases, *J*-band spectroscopy appears an attractive approach for direct metallicity determinations of red stars out to large distances (B. Davies et al. 2010; Evans et al. 2011; Gazak et al. 2014, see Fig. 5). Moreover, with AO-corrected IFUs multiple stars can be observed simultaneously per IFU (leading to a large effective multiplex compared to the number of IFUs).

Observations with both modes will be highly complementary, probing different regions and populations (thus providing good sampling of each spatial and kinematic feature). From sensitivity calculations (e.g. Evans et al. 2011; v2 of the White Paper), MOSAIC spectroscopy will open-up a huge new number of galaxies for detailed study for the first time, with the capability to determine stellar metallicities out to Mpc distances (for stars near the tip of the RGB) and, in the case of the AO-corrected *J*-band observations, out to tens of Mpc for red supergiants.

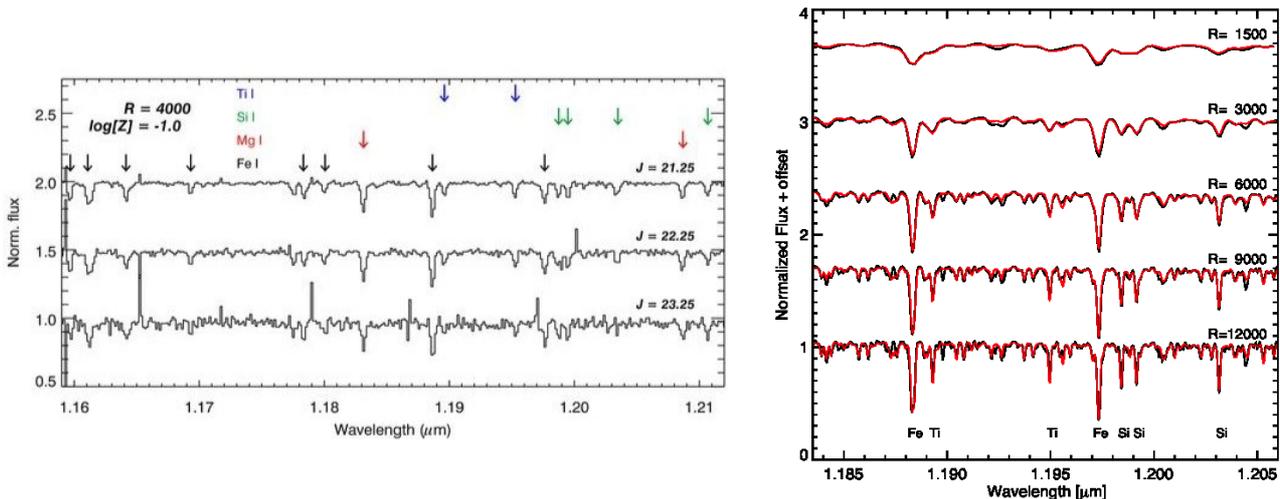

**Figure 5:** *Left:* Simulated *J*-band spectra of red supergiants to investigate the potential of this spectral region for metallicity determinations (Evans et al. 2011). *Right:* Native (*R*=12000) and degraded Subaru-IRCS observations of a Galactic red supergiant (and model fits, shown in red) used to determine that *R*>3000 is required to recover accurate metallicities (Gazak et al. 2014).

## 2.6. SC6: Galaxy Archaeology

Significant effort has been invested over the past few decades in searching for primordial stars in the Galaxy, which are the long-lived descendants of the earliest stellar generations. These stars will have formed from a (near-)pristine ISM, which will have only been weakly enriched in metals from the first supernovae. Their atmospheres therefore give us a fossil record of the ISM from redshifts of $z \geq 10$. Having a direct tracer of chemical abundances at such an early time can provide fundamental constraints on the properties of the first generations of star formation, as well as giving indirect information on their masses and ionising feedback, of interest in the context of reionisation.

A key question is whether the galaxies in the Local Group formed from the early primordial gas or from gas that had already been partially enriched, giving a metallicity 'floor'. This has important cosmological implications. In the standard hierarchical scenario the first structures to form are dwarf galaxies which subsequently merge to form larger structures like the Galaxy and other massive disk galaxies. If the metallicity distribution functions (MDFs) of Local Group dwarfs display such a floor, then either the hierarchical galaxy formation model is wrong, or the present-day 'surviving' dwarf galaxies formed later (in which scenario, they are not the relics of the primordial dwarfs).

The Hamburg-ESO Survey (Christlieb et al. 2008) revealed an apparent floor of [M/H] = –3.5 (Schörck et al. 2009), in agreement with theoretical predictions, but the discovery of an extremely metal-poor star which is not strongly enhanced in C or O has challenged this model (Caffau et al. 2011). Major progress in our understanding of the metal-poor MDF can be expected in the coming years from, e.g., the Sloan Digital Sky Survey (SDSS, York et al. 2000) and the SkyMapper telescope (Keller et al. 2007). For instance, the dramatic finding of a star with [Fe/H] < –7.1 (Keller et al. 2014) and new larger samples of very metal-poor stars (Norris et al. 2013; Cohen et al. 2013; Roederer et al. 2014).

Such efforts in the Galaxy are an important foundation, but to provide rigorous tests of galaxy evolution models we need similar observational constraints for the dwarf (e.g. the Fornax Dwarf, shown in Fig. 6) and in the newly-discovered, ultra-faint dwarf galaxies of the Local Group. In extragalactic studies we are currently limited to spectroscopy of giant stars, which yield insufficient samples for firm statistics in the metal-poor tail of their MDFs (i.e., if the fraction of stars below [Fe/H] ~ –4 is $10^{-4}$, then at least $10^4$ stars need to be observed to find just one!) Thus, we need to observe dwarfs and stars near the main-sequence turn-off, which requires the sensitivity of the E-ELT.

Similarly, the formation of the Galactic bulge – probably the oldest component of the Galaxy – remains controversial. Its stellar kinematics are compatible with those of an old spheroid, whereas those for the more metal-rich population are compatible with a bar, suggesting differing formation scenarios (Babusiaux et al. 2010). Aside from observations of a few tens of dwarfs via microlensing techniques (e.g. Bensby et al. 2013), observations in the bulge are also limited to giant stars at present and only with MOSAIC will we be able to disentangle its complex mix of stellar populations.

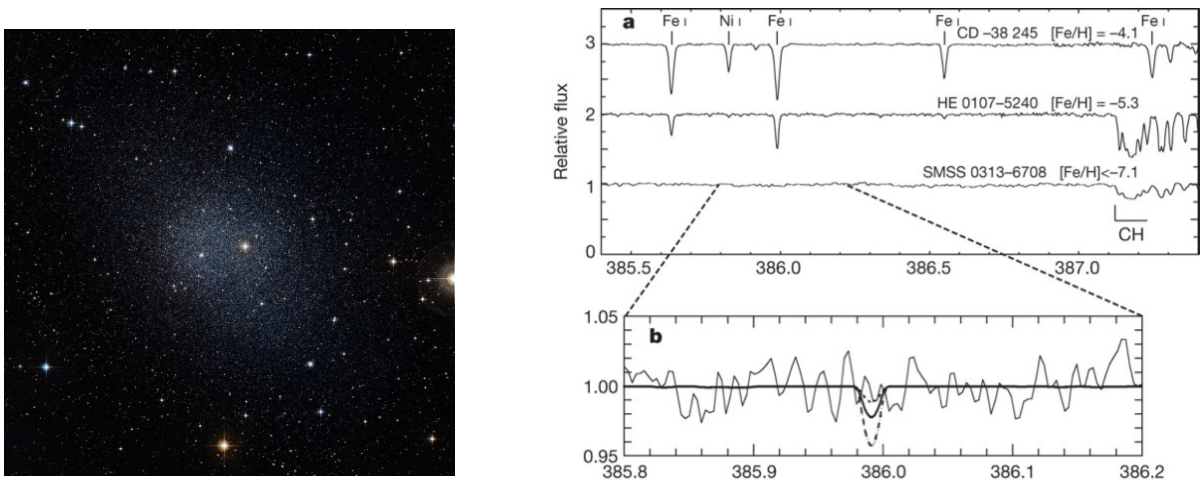

**Figure 6:** *Left:* Main-sequence turn-off stars in systems such as the Fornax Dwarf hold vital clues to the history of the Local Group but are out-of-reach of current spectrographs. *Right:* The exceptionally metal-poor star discovered by Keller et al. (2014), compared to other metal-poor stars; precise abundances for such stars require high-resolution spectroscopy.

## 2.7. SC7: Galactic Centre Science

One of the most compelling astrophysical laboratories available to us is the central parsec of our own Galaxy, providing a unique view of the processes at the heart of a massive galaxy. Indeed, one of the most spectacular results of the past decade was arguably the observed orbits of stars around Sgr A*, the central black-hole (Ghez et al. 2005; Eisenhauer et al. 2005). Studies of these 'S' stars in the central arcsec (~0.04 pc) have since been used to provide the most accurate estimates of the distance to the GC to date (8 kpc) and the mass of the central compact object ($4\times10^6$ $M_\odot$). However, studies in this region remain challenging even with 8-m class facilities, primarily because the line-of-sight extinction is so high ($A_V = 30$, $A_K = 3$). We now briefly consider the contribution of the E-ELT to GC studies, highlighting cases where we are currently limited by sensitivity rather than spatial resolution.

*The parsec-scale central stellar cluster:* Beyond the central arcsec there are ~100 known massive (OB-/WR-type) stars in the inner 0.5 pc, with debate as to whether they are structured in one or two disks (e.g., Lu et al. 2009 cf. Paumard et al. 2006). Two scenarios have been proposed for the formation of these stars: in-situ, in an accretion disk orbiting Sgr A*, or in a massive star-cluster which migrated onto Sgr A* and dissolved in the central pc. E-ELT spectroscopy of the fainter (lower-mass) stars will provide the large sample required to investigate the origins of this important cluster. While HARMONI will provide spectroscopy of the innermost region, MOSAIC observations of the stars at larger radii will be a vital component as the two formation models predict different surface density and mass profiles.

*Gas content of the regions around Sgr A*:* Moving further out, the clumpy, torus-like Circumnuclear Disk (CND, with inner and outer radii of ~2 and 7 pc, respectively) is also of interest as it has been suggested as a potential stellar nursery. The hole within the CND is occupied by the 'Minispiral', an ensemble of gas and dust clouds (Sgr A West), each of which orbit Sgr A* in their own planes (and the northern arm of which has been suggested as a clump of the CND which has fallen inwards after a shock of some sort). Although it appears quiescent in terms of current star formation, the CND is thought to be an important ingredient of the GC region – it may later evolve into a real accretion disk, unleashing an AGN-like event in the Milky Way and forming a new disc of stars. Several groups are currently looking for star-formation sites or self-gravitating nodes in the CND, and resolving its dynamics would provide important information on its properties and likely future. Near-IR spectroscopy of the region with the ELT will enable simultaneous detections of stars and molecular ($H_2$) and ionised (HI) gas, from which we can study their respective interactions. An ELT-MOS programme with IFUs could potentially map the large majority of the neutral-ionised CND region (see Fig. 7). Assuming a field of ~2"x2" for each IFU, this would require >>100 HARMONI pointings but, depending on the multiplex, MOS observations would be *significantly* faster.

*Stellar populations in the inner Galaxy:* The stellar populations beyond the central region in Fig. 7 comprise massive star clusters (e.g. the Arches and Quintuplet clusters), apparently isolated massive-stars, other gas clouds, and X-ray sources (Mauerhan et al. 2010). For the large majority of the objects in these high-mass populations, good signal-to-noise spectroscopy requires the sensitivity of the E-ELT to provide insights into their physical properties, binarity, and dynamics. Moreover, the *Chandra* observations from Mauerhan et al. detected >6000 X-ray sources, so a thorough and efficient census of these objects to determine their natures would require a high-multiplex ELT-MOS.

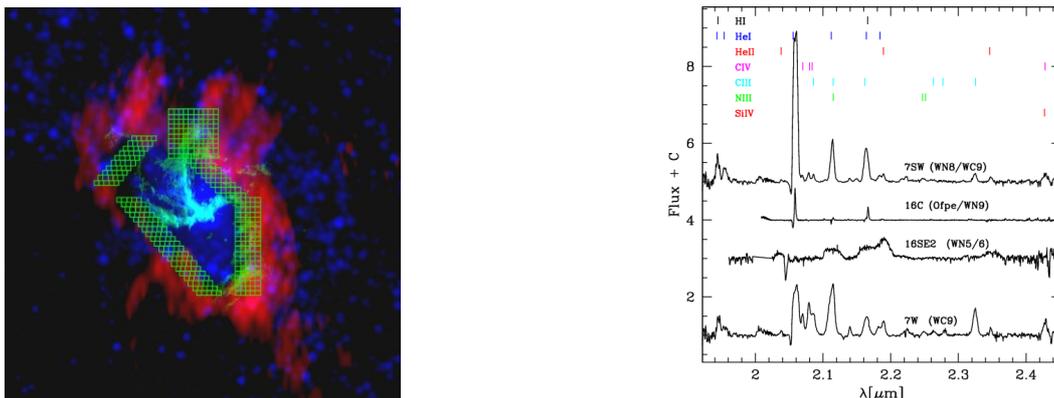

**Figure 7:** *Left:* Example IFU fields (2"x2") to map the CND/Minispiral interface. (Image credits: NRAO/AUI; Yusef-Zadeh, Wright & Stolovy.) *Right:* Only a small number of the most luminous emission-line stars in the central cluster are accessible with current facilities (e.g. with VLT-SINFONI, Martins et al. 2007); spectroscopy of the less-luminous/-massive populations requires the E-ELT.

## 2.8. SC8: Planet Formation in Different Environments

The most significant addition to the previous ELT-MOS cases is new work concerning studies of exoplanets in stellar clusters and, potentially, in external galaxies. The two most commonly discussed formation mechanisms for planets are:

- *Core accretion:* A solid core forms over a time-scale of ≥1 Myr and, if massive enough (5-10 $M_{Earth}$), hydrogen and helium is accreted from the circumstellar disk, ultimately leading to the formation of a gas giant planet.
- *Disk instability:* Alternatively, planets may form directly via an instability of the disk. While this process is very fast, it requires a massive disk to work.

The task ahead is to understand which physical parameters are most relevant to the formation process(es). For example, if planets form quickly via disk instability, the mass-ratio of the star to the disk would be important, but the lifetime of the disk would be less relevant. In contrast, the lifetime of the disk and the abundance of heavy elements are both important for the core-accretion scenario.

Observationally, the probability for a star to have a massive planet appears to depend on both the stellar mass and metallicity, with the latter point taken as evidence for the core-accretion process. However, recent ALMA observations of proto-planetary disks have shown that they are complicated structures with local density enhancements and regions with varying sizes of dust grains (e.g. van der Marel et al. 2013). Moreover, radial variations in the silicate chemistry in proto-planetary disks (specifically the olivine to pyroxene ratio) were found by VLTI observations (van Boekel et al. 2004). These results suggest that the detailed structures and chemistry of the disks appear to be important factors in planet formation as well as their global properties. In addition, external factors are also thought to influence planet formation, and we now briefly discuss two environmental effects.

### 2.8.1. Effects of radiative feedback and stellar density

It is thought that ~50% of stars form in clusters/associations with ≥ ~1750 stars. In the context of planet formation, UV radiation from the most massive star(s) in a cluster can photo-evaporate circumstellar disks (e.g. Fig. 8), such that their expected lifetimes are only $10^5$-$10^6$ yrs. This UV radiation could therefore strongly influence planet formation in stellar clusters (e.g., Armitage, 2000; Fatuzzo & Adams 2008; Holden et al. 2010), but does it act to inhibit it?

Interestingly, recent results suggest that the Sun formed in a cluster with ~1200 stars (Gounelle & Meynet, 2012; Pfalzner et al. 2013), arguing that our solar system must have formed in a clustered environment. Equally, we do not see evidence for low-mass stars forming first in clusters/associations (e.g. Preibisch & Zinnecker, 2007), i.e., the massive stars did not appear to form after potential planet formation. Lastly, and perhaps counterintuitively, photo-evaporation of the disks may actually help rather than inhibit planet formation (Throop & Bally, 2005; Mitchell & Stewart, 2010). In this scenario the UV radiation of nearby stars may trigger planet formation because an increasing dust-to-gas ratio arises in the shielded interior, such that the dust becomes gravitationally unstable.

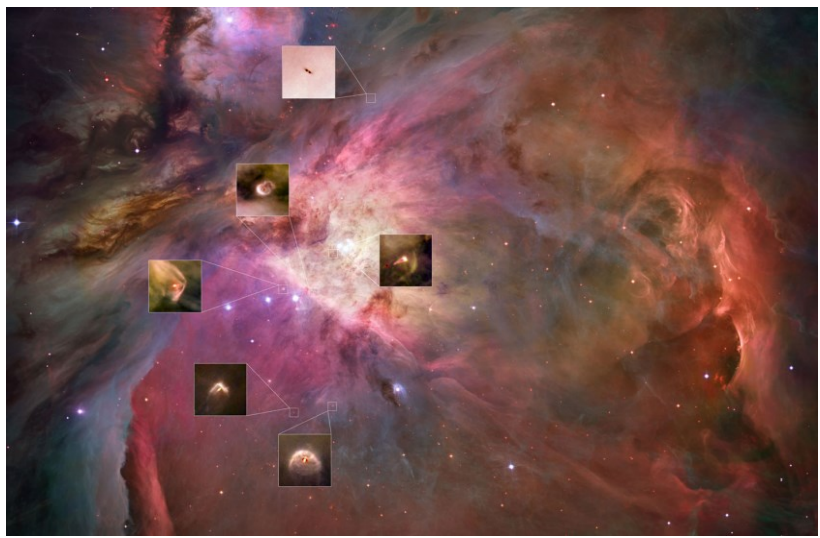

**Figure 8:** Evaporation of proto-planetary disks in Orion; most planets likely formed in such an environment.

These issues are still unresolved at present, although Thompson (2013) recently predicted that in *very* dense systems (nuclear star clusters, starburst clusters, compact high-$z$ galaxies, some Galactic globular clusters) giant planets should not form as the UV radiation would lead to disk temperatures in excess of the ice-line (~150-170 K). This poses some interesting observational tests as we look ahead to ELT observations. For instance, searches for planetary transits in 47 Tuc have so far revealed no detections (Gilliland et al. 2000; Weldrake et al. 2005). The globular cluster NGC 6366 is less dense than 47 Tuc but has a comparable metallicity ([Fe/H] = –0.82) such that Thompson predicted we should expect detections of giant planets there if density is more important than metallicity. He also argued that *metal-rich* globulars (e.g. NGC 6440) should be devoid of giant planets due to their comparable stellar densities to 47 Tuc.

### 2.8.2. Dynamical ejection

The density of the star-forming region can play a further role in planet formation (and evolution) via close encounters of stars and circumstellar disks. For instance, Bate et al. (2010) predicted the truncation (or even destruction) of proto-planetary disks by dynamical interactions between stars, while in dense globular clusters, Bonnell et al. (2011) predicted that planets with orbits of >0.1 AU could be ejected from their stellar systems. (In contrast, in open clusters only the wider-orbit systems are disrupted and the impact on planet formation is much less significant.) More recent simulations by Hao et al. (2013) have considered the fraction of planetary survivors at different radii for a planetary system like our own solar system in the environment of the Orion Nebula Cluster – as one might expect the inner planets have a higher probability of surviving stellar encounters, while at > ~10 AU the outer planets can more easily escape the system. Hao et al. (2013) also concluded that planets such as Saturn, Neptune, and Uranus can be affected by both encounters with stars and planet-planet scattering, while a Jupiter-like planet is only influenced by stellar encounters given its mass.

### 2.8.3. Planets in Stellar Clusters – Initial Results

While environment appears to play an important role in planet formation/evolution, efforts to detect planets in stellar clusters over the past decade have had somewhat mixed results, as follows:

- *Hyades cluster:* Expected to have a high incidence of planets given its high metallicity, initial radial velocity (RV) surveys did not find any (Cochran et al. 2002; Guenther et al. 2005). A planet (with M = 7.6±0.2 $M_{Jup}$, P = 594.9±5.3 days) was later discovered around the intermediate-mass star ε Tau (Sato et al. 2007). More recently, Farihi et al. (2013) have inferred rocky planetesimals to be in orbit around two white dwarfs.

- *M67:* A first RV survey of this open cluster did not detect any planets (Pasquini et al. 2012), but three planets have recently been discovered from observations over a longer baseline (Brucalassi et al. 2014).

- *NGC 6252 & NGC 6791:* No detections to date in these metal-rich open clusters (Montalto et al. 2007; 2011).

- *NGC 2423 & NGC 4349:* Two substellar objects (M = 10.6 and 19.8 $M_{Jup}$) were found orbiting intermediate-mass stars in these intermediate-age open clusters (Lovis & Mayor, 2007).

- *Praesepe cluster (M44):* Two planets were found from RV monitoring of 53 stars, giving a lower limit of $3.8_{-2.4}^{+5.0}$ % on the hot Jupiter frequency in this metal-rich open cluster (Quinn et al. 2012).

- *NGC 6811:* Two Neptune-sized planets were discovered in this open cluster by Meibom et al. (2013), who argued that the frequency and properties of planets in open clusters appear consistent with those for Galactic field stars (which follows logically if most of the field stars initially formed in clusters).

- *Globular clusters:* No planets have been detected in 47 Tuc from transit-monitoring campaigns with both the *HST* and ground-based telescopes (Gilliland et al. 2000; Weldrake et al. 2005); similarly for ground-based monitoring in ω Cen (Weldrake et al. 2008). This led Gilliland et al. to argue that planets in 47 Tuc must be ten times rarer than for Galactic field stars. Indeed, there is only one known planetary system in a globular cluster to date – a hierarchical triple system with a Jupiter-mass planet in a wide orbit with a binary millisecond pulsar (PSR B1620-26; Backer 1993; Thorsett et al. 1999).

In summary, globular clusters appear to have a much lower frequency of planets than stars in the Galactic field population or in open clusters. Interestingly, there does not appear to be a strong correlation between planet frequency and metallicity for the open clusters, in contrast to what one might expect if the relation between metallicity and planet formation was the defining characteristic; this could be a first hint that stellar density is the more important influence.

### 2.8.4. Example Observing Programmes for MOSAIC

MOSAIC will provide comprehensive studies of main-sequence and giant stars in a much broader range of clusters than possible with current facilities, e.g., in both open and globular clusters, spanning a range of densities and metallicities. There will be strong synergies with the clusters component of the *Gaia*-ESO Spectroscopic Survey which will provide quantitative physical parameters for cluster members in the coming years (vital for subsequent RV analysis), as well as identifying binaries to be excluded from further monitoring. To illustrate the expected RV signals, the velocity semi-amplitudes of known planet-hosting giant stars are typically larger than 30 to 40 m/s, as shown in Fig. 9.

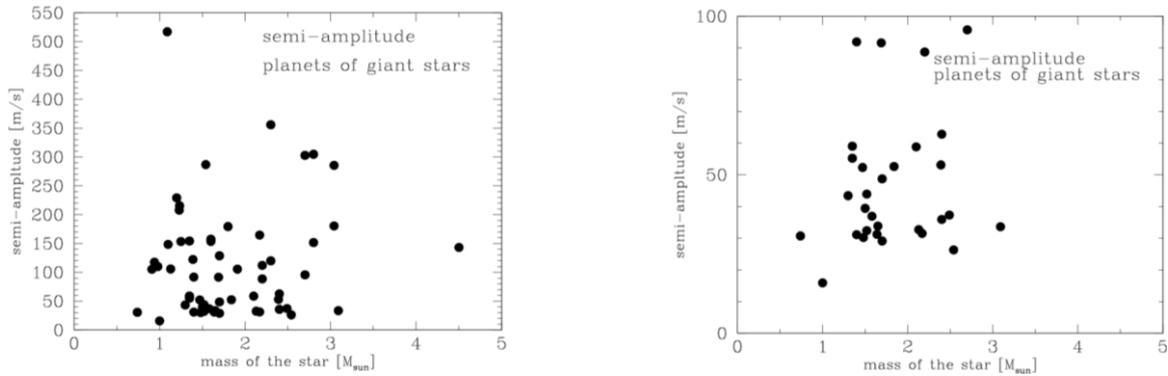

**Figure 9:** *Left:* Radial-velocity (RV) semi-amplitudes of known planet-hosting giant stars. These massive planets are relatively easy to detect compared to lower-mass planets around lower-mass stars. *Right:* Expanded view for lower RV semi-amplitudes – almost all the known planet-hosting giant stars have variations of $\Delta RV \geq 30$ m/s.

Example MOSAIC programmes in this area include:

- *Globular clusters:* As noted earlier, observations in globulars will provide important tests of the relative roles of stellar metallicity and density on the frequency of planets (e.g. via campaigns in 47 Tuc, NGC 6366, NGC 6440, etc). Main-sequence G-type stars in 47 Tuc have $V = 18$ mag, while K-type giants have $V = 15$ mag.

- *Planetary frequency vs. Galactocentric distance:* Due to the metallicity gradient of the Milky Way, we expect that the frequency of planets should decrease with increasing Galactocentric distance. For example, the frequency of planets at 7 kpc should be twice that at 9 kpc (Reid, 2006). Thus, we would like to be able to detect planets around stars up to ~1 kpc from the Sun, which entails observations of main-sequence stars with $V = 12$ mag. A more challenging goal is to search for planets around stars in the Galactic bulge where, with the larger distance and much greater extinction, giant stars have $V = 17$ mag.

- *Planets in Local Group galaxies:* Further afield, we can also ask if exoplanets exist in other galaxies and, if so, whether they are similar to those found in the Galaxy. A first step would be to obtain precise RVs for 'hot Jupiters' around stars in Local Group galaxies. For example, giant stars in the Sagittarius Dwarf Elliptical Galaxy (Sag. DEG, at ~20 kpc) have $V = 18$ to 20 mag, and are potentially within the grasp of RV studies with the E-ELT. The Sag. DEG has a considerably lower stellar density than the solar neighbourhood, and with a range of metallicities ($-1.5 \leq [Fe/H] \leq 0.0$), so this would extend planet searches into a very different environment compared our own Galaxy.

A common requirement for these cases is to detect a Jupiter-mass planet with an orbital period of P < 10 (100) days which, for a solar-type star would cause RV variations with a semi-amplitude of 94 (44) m/s. Therefore the prime requirement for detection sensitivities is to be able to detect planets with semi-amplitudes of $\geq 40$ m/s for main-sequence stars with $V \leq 18$ mag, and for giant stars with $V \leq 15$ mag.

### 2.8.5. Astrophysical Limitations

A range of issues can impact on RV observations for planet detection, including:

*Star spots:* RV periodicity can potentially be caused by spots on the stellar surface. Thus, if candidate planets are found we should check if the variations are comparable to the stellar rotation period (requiring photometric data with a medium-sized telescope) and, while the signal caused by planets will always be the same, the properties of any spots will change with time. Additional checks would include determination of RVs for lines with different temperature sensitivities (to look for differences from the two sets of lines which might indicate their origin from a spot) and looking for indications of stellar activity (e.g. the Ca H+K lines, changes in line widths).

*Contaminating stars:* A faint star contained within the PSF of a target star can also mimic a planet so a first check will be to obtain high-resolution, AO-corrected images of any candidates (for which the angular resolution of the VLT is sufficient). Beyond such initial checks and assuming that our main targets are K-type giants, contamination by an early-type star (with fewer spectral lines) will not significantly impact on the RV estimates. For a contaminating later-type star, 2.5 mag fainter than the primary target and with $\Delta RV \geq 20$ km/s, we find from simulations that we expect a shift of < 3 m/s for the target, and the two stars also produce well-separated peaks in the cross-correlation function so, in the context of searching for giant planets, the RV results will be reliable.

*Stellar oscillations:* The periods arising from stellar oscillations are typically of order a few minutes for solar-like stars and a few hours for giants. Such short orbital periods would be impossible for planets around such stars, so oscillations should not lead to any misidentifications. However, RV-variations from oscillations will be an important source of jitter noise in planet-search observations. The oscillation amplitudes are much larger for giants than dwarfs, as they scale with both luminosity and mass; the maximum frequency is given by Kjeldsen & Bedding (1995). As an example, β Gem is a planet-hosting K0-type giant (M = 1.7±0.4 $M_\odot$), for which seven (stellar) oscillation modes have been identified (Hatzes et al. 2006; Hatzes & Zechmeister 2007). These cause an RV-jitter of 21 m/s (rms) and the largest mode has an amplitude of 5 m/s. In contrast, the semi-amplitude of a Jupiter-like planet orbiting a solar-like star at 1 AU is 28 m/s, i.e. stellar oscillations – at least for giants – should not limit planet detection.

### 2.8.6. Performance Simulations

To investigate the potential of detecting giant planets with MOSAIC we undertook tests with a range of synthetic spectra. Using the current E-ELT exposure time calculator we estimated that S/N ~100 can be obtained for stars with $V$ = 18 mag in an exposure time of only 300s (assuming a K-type star and an instrument efficiency of 25%); for $V$ = 19 and 20 mag the exposure times increase to 700 and 2200s, respectively. While these calculations were for the previous telescope design (with a 42m primary aperture), they highlight the feasibility of such observations.

The RV accuracy (δRV, in m/s) that can be achieved for solar-like stars was given by Hatzes & Cochran (1992):

$$\delta RV \approx 1.45 \times 10^9 \times (S/N)^{-1} R^{-1.5} B^{-0.5},$$

where B is the wavelength coverage of the observed spectral range (in Å). To investigate the RV accuracy which can be recovered we generated simulated spectra for a range of stellar effective temperatures ($T_{eff}$), logarithmic gravities (log $g$), and metallicity ([Fe/H]), and for S/N = 100 and 20 (e.g. Fig 10). The simulations employed ten runs of synthetic spectra (with appropriate random noise characteristics) and assume a wavelength range of 500 to 600 nm (i.e., B = 100 nm in the above equation), $R$ = 20,000, and 3-pixel sampling. These simulated spectra were then analysed using the same cross-correlation methods as for real observations; the limiting RV results are summarised in Table 2. Compared to dwarfs, greater accuracy can be achieved for giant stars due to their typically stronger absorption lines, while less accurate RVs will be recovered for metal-poor stars due to the weaker lines (see Fig. 10).

**Table 2:** Summary of predicted RV accuracy (stellar jitter) for a range of physical stellar parameters (for S/N = 100 and 20).

| $T_{eff}$ | log $g$ | [Fe/H] | RV accuracy [m/s] | |
|---|---|---|---|---|
| | | | S/N=100 | S/N=20 |
| 6000 | 4.0 | 0.0 | 11 | 57 |
| 6000 | 4.0 | −0.5 | 13 | 73 |
| 6000 | 4.0 | −1.5 | 29 | 170 |
| 5500 | 4.5 | 0.0 | 10 | 54 |
| 5500 | 4.5 | −0.5 | 15 | 41 |
| 5500 | 4.5 | −1.5 | 28 | 78 |
| 5000 | 2.5 | 0.0 | 8 | 40 |
| 5000 | 2.5 | −0.5 | 8 | 46 |
| 5000 | 2.5 | −1.5 | 15 | 69 |
| 4000 | 1.5 | 0.0 | 4 | 30 |
| 4000 | 1.5 | −0.5 | 9 | 64 |
| 4000 | 1.5 | −1.5 | 10 | 25 |

*Note:* Reducing the wavelength coverage from 100 to 40 nm should, in principle, decrease the RV accuracy by a factor of 1.7 but, by selecting a spectral range with numerous deep, narrow lines (~500-540 nm), it only decreases it by a factor of 1.2.

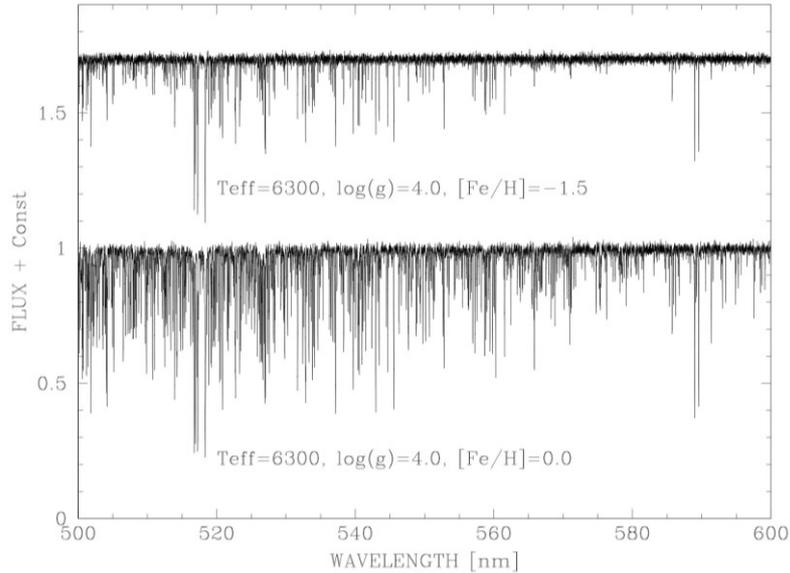

**Figure 10:** Simulated dwarf spectra with identical effective temperatures (Teff) and gravities (log *g*), but different metallicities. We can recover more precise radial-velocity (RV) estimates for the metal-rich spectrum given the greater intensity/density of absorption lines. Similarly, the 500-550 nm region is richer in spectral features than 550-600 nm.

### 2.8.7. Number of spectra required per field

To detect a planet with a RV semi-amplitude that is twice the noise level, with a false-alarm probability of 1% (50%), requires 80 (30) epochs (Cumming et al. 2004). If the semi-amplitude is four times the noise only 28 epochs are required to reduce the false-alarm probability to 1% and, with additional measurements, the probability rapidly decreases (e.g. to 0.01% with 35 epochs). Thus, with ~30 epochs we can reliably detect planets with RV semi-amplitudes of 40 m/s, corresponding to a Jupiter-mass planet with a period of 130 days around a solar-like star (Fig. 11).

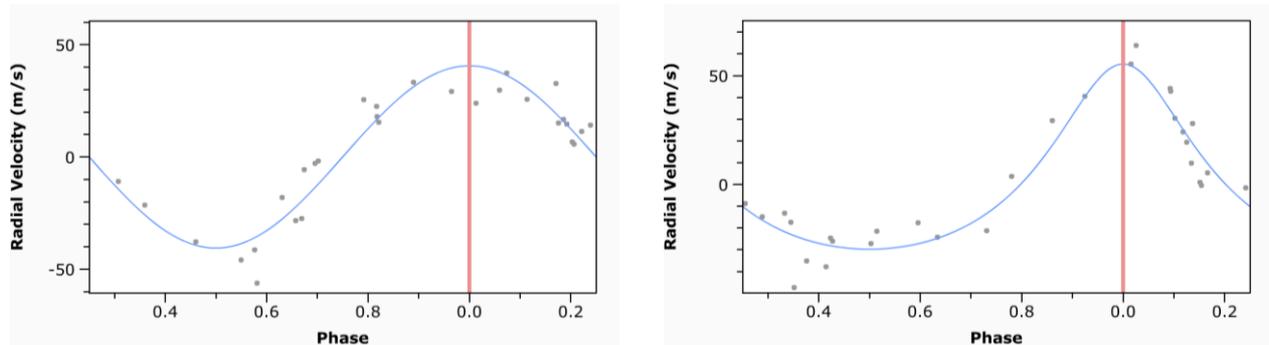

**Figure 11:** Simulated RV-data for 30 observations (with δRV = 10 m/s) of a Jupiter-mass planet orbiting a solar-like star (P = 130 days). *Left:* Circular orbit. *Right:* Eccentric orbit (ε = 0.3) for longitude = 0 deg. (Calculated using the University of Nebraska-Lincoln Radial Velocity Simulator.)

### 2.8.8. Time-series Spectrophotometry

A further exoplanet case concerns the potential use of IFU spectroscopy to investigate the atmospheric properties of transiting planets and free-floating brown dwarfs. For examples of what is possible with current instrumentation:

- Crossfield et al. (2013) measured the radius of the warm ice giant GJ 3407b in six narrow bands from 2.09-2.36 μm using Keck-MOSFIRE, and could rule out both cloud-free or moderately metal-enriched atmospheric models from their results.

- Biller et al. (2013) used the MPG-ESO 2.2m GROND multi-band imager to obtain simultaneous six-band (*r'i'z'JHK*) photometric time series for the two closest known brown dwarfs, in the Luhman 16AB binary system (Luhman, 2013). The B component was known to display variability with a ~5 hour period (Gillon et al. 2013), and Biller et al. found 1-15% variability across multiple bands for both components, including phase offsets between bands for the B component. These phase offsets correlate with the model atmospheric pressure probed at each band, in other words, they reveal complex vertical cloud structures driving the variability.

Looking to the future, an AO-fed multi-IFU instrument would enable high-precision spectroscopic time-series work for directly-imaged exoplanet companions, as well as fainter brown dwarfs and transiting planets. To date, ground-based spectrophotometric time-series studies have been hindered by two factors: 1) instability due to changing telluric water absorption and 2) slit losses with changing parallactic angles, which prevent accurate photometric calibration across time series. The use of multi-IFU instruments directly solves both these problems: changing telluric water absorption can be tracked by observing multiple reference stars on different IFU arms, and a slitless IFU design would minimise slit losses. Pilot multi-IFU studies are currently underway: Biller and collaborators have recently acquired 7.5 hours of time-series data with VLT-KMOS. Multi-IFU studies are only possible at present for comparatively bright objects, so a multi-IFU mode for MOSAIC would enable such studies for much fainter objects. In particular, the inclusion of AO should boost PSF-stability and also allow time-series work for close companions.

## 3. MOSAIC TOP-LEVEL REQUIREMENTS

The instrument requirements that flow down from each SC are summarised in Table 3 (in which the estimated multiplex for each SC assumed a 40 arcmin$^2$ patrol field); where relevant, the requirements for the 'high definition' and 'high multiplex' modes are captured. The details and specific motivations for these requirements will be expanded on in v2 of the White Paper. In particular, we comment that the best aperture for the single-object mode is that which maximises the S/N ratio for point-line sources with GLAO correction; given the expected performances this equates to ~0.6 arcsec (or 0.8 arcsec for seeing-limited observations). As mentioned in SC2, note that the requirements for the IGM case are slightly different to the others as they require *optical* IFU observations (but without strong requirements on the spatial resolution, i.e. GLAO is sufficient).

**Table 3:** Summary of top-level requirements from each Science Case; 'desirable' requirements are shown in italics.

| Case | Multiplex | FoV/target | Spatial pixel size | λ-coverage (μm) | *R* |
|---|---|---|---|---|---|
| SC1<br>*First light* | 20-40 | 2" × 2" | 40-100 mas | 1.0-1.8<br>*1.0-2.45* | 5,000 |
| | ≳150 | – | (GLAO – 0.6"∅) | 1.0-1.8<br>*1.0-2.45* | >3,000 |
| SC2<br>*IGM &*<br>*Gal. clusters* | 10-15 | 2" × 2" | (GLAO – IFU) | 0.4-1.0<br>*0.37-1.0* | >3,000 |
| | 50-100 | – | (GLAO) | 0.6-1.8<br>*0.6-2.45* | >3,000 |
| SC3<br>*Gal evol.* | ≳10 | 2" × 2" | 50-80 mas | 1.0-1.8<br>*1.0-2.45* | 5,000 |
| | ≳100 | – | (GLAO – 0.6"∅) | 1.0-1.7<br>*0.8-2.45* | ≥5,000<br>*~10,000* |
| SC4<br>*AGN* | ~10 per field | 2" × 2" | ≤100 mas | 1.0-1.8 | >3,000 |
| SC5<br>*Extragal*<br>*stellar pops.* | Dense | 1" × 1"<br>*1.5" × 1.5"* | ≤75 mas<br>*20-40 mas* | 1.0-1.8<br>*0.8-1.8* | 5,000 |
| | 10s arcmin$^{-2}$ | – | (GLAO) | 0.4-1.0 | ≥5,000<br>*≥10,000* |
| SC6<br>*Gal archaeol.* | 10s arcmin$^{-2}$ | – | (GLAO) | 0.41-0.46 & 0.64-0.68<br>*0.38-0.52 & 0.60-0.68* | ≥15,000<br>*≥20,000* |
| SC7<br>*GC science* | Dense | ≥2" x 2" | ≤100 mas | 1.5-2.45 | ≥5,000<br>*≥10,000* |
| SC8<br>*Planet form.* | 10s per field | – | (GLAO) | 0.5-0.6 | ≥20,000 |

*Note:* Minimum target size for SC1 is reduced to 1"x1" if on/off sky subtraction is used.

## 3.1. Simultaneous high-definition/high-multiplex observations

If the proposed high-definition and (or) high-multiplex (IGM IFU) modes were to have independent spectrographs, parallel observations with two modes could significantly boost the operational efficiency of the E-ELT. For instance, multi-wavelength surveys have (necessarily) concentrated on a small number of deep fields for studies of distant galaxies, e.g., the *Chandra* Deep Field South, the COSMOS field, etc. Given the huge investment of resources in these deep fields, the E-ELT will almost certainly be used to observe high-$z$ galaxies located within them. Moreover, even if new regions are observed to a comparable depth (and breadth of wavelengths) in the coming years, they will still be relatively limited in number. Thus, many of the high-$z$ targets envisaged for the cases mentioned in SC2 and SC4 are in the same regions of the sky. The same is also true for the stellar populations cases, where there are common galaxies of interest. In such a scenario, even a small number of `Large Programmes' executed in parallel (of a couple of hundred hours each), could potentially save operational costs of order M€s (i.e. becoming comparable to meaningful shares of the individual instrument budgets).

Operational issues such as exposure times, scheduling, and sky subtraction will need to be explored in greater detail in the next phase, but do not initially appear to preclude parallel observations. Equally, the different image-quality requirements (GLAO/MOAO) should not be a problem for such an instrument. Simulations from Basden et al. (2012) demonstrated that the image quality in regions 'between' MOAO-corrected sub-fields is not significantly diminished compared to the stand-alone GLAO correction. Indeed, they concluded that they might actually be able to improve on the GLAO correction in some directions, without impacting on the performance for the MOAO-corrected sub-fields.

## 3.2. A link to a high-resolution (HIRES) spectrograph

The two modes of MOS observations discussed (high multiplex, high definition) likely imply different 'pick-offs' to select targets in the E-ELT focal plane, analogous to the single-object and IFU modes which feed the VLT-FLAMES Giraffe spectrograph. Within such an architecture one could envisage a fibre-link/relay to feed a high-resolution (HIRES) spectrograph to enable parallel observations (as per the fibre-feed to UVES from FLAMES). For example, deep MOS observations of faint IGM sight-lines to LBGs, could be complemented with simultaneous HIRES observations of a sight-line to a bright quasar.

## 4. SUMMARY

The E-ELT will be the world's largest optical/IR telescope when complete, so we will aim for the largest possible discovery space when designing MOSAIC, balanced by technical feasibility and cost. In this contribution we have highlighted the range of MOS cases compiled with input from the astronomical community at workshops, large conferences, and at smaller meetings across Europe and in Brazil. To deliver the observations required by these cases we have defined two primary modes – high multiplex and high definition – and the majority of the cases described here would benefit from the combination of these two modes. We note that a visible/near-IR MOS with such capabilities is technically feasible, as demonstrated by recent studies of critical issues such as sky-background subtraction and MOAO.

In the coming year we will prioritise the science requirements and take into account both technical and operational feasibilities as part of a proposed Phase A study of the MOSAIC concept. In the study we will aim to deliver a versatile and very capable MOS to the E-ELT as soon as possible in its operations – this will enable a broad and exciting science programme, drawn from the cases summarised in this contribution and beyond.

# REFERENCES

Armitage, 2000, A&A 362, 96
Arnaboldi et al. 2002, AJ, 123, 760
Babusiaux et al. 2010, A&A, 519, 77
Backer, 1993, in *Planets Around Pulsars*, 36, 11
Basden et al. 2012, SPIE, 8447, 5E
Bate, Lodato & Pringle, 2010, MNRAS, 401, 1505
Battaglia et al. 2008, MNRAS, 383, 183
Bensby et al. 2013, A&A, 549, A147
Biller et al. 2013, ApJ, 778, L10
Bonnell et al. 2001, MNRAS, 322, 859
Bouwens et al. 2007, ApJ, 670, 928
Brucalassi et la. 2014, A&A, 561, L9
Burke et al. 2012, MNRAS, 425, 2058
Caffau et al. 2011, Nature 477, 67
Cano-Díaz et al. 2012, A&A, 537, L8
Capak et al. 2011, Nature, 470, 233
Carrera et al. 2007, AJ, 134, 1298
Christlieb et al. 2008 A&A, 484, 721
Cochran et al. 2002, AJ, 124, 565
Cohen et al. 2013, ApJ, 778, 56
Cole et al. 2004, MNRAS, 347, 367
Crossfield et al. 2013, A&A, 559, 33
Croton et al. 2006, MNRAS, 365, 11
Cuby et al. 2010, SPIE, 7735, 80
Cumming, 2004, MNRAS, 354, 1165
Davies, B. et al. 2010, MNRAS, 407, 1203
Davies, R. et al. 2010, SPIE, 7735, 77
Eisenhauer et al. 2005, ApJ, 628, 246
Evans et al. 2011, A&A, 527, A50
Evans et al. 2012, SPIE, 8446, 7K
Evans et al. 2013, '*ELT-MOS White Paper*', arXiv:1303.0029
Farihi et al. 2013, MNRAS, 432, 1955
Fatuzzo & Adams, 2008, ApJ, 675, 1361
Gadotti & Sánchez-Janssen, 2012, MNRAS, 423, 877
Gazak et al. 2014, ApJ, 788, 58
Ghez et al. 2005, ApJ, 620, 744
Giallongo et al. 2014, ApJ, 781, 24
Gilliland et al. 2000, ApJ, 545, L47
Gillon et al. 2013, A&A, 555, L5
Glazebrook, 2013, PASA, 30, 56
Gobat et al. 2013, ApJ, 776, 9
Gounelle & Meynet, 2012, A&A, 545, A4
Guennou et al. 2012, A&A, 537, 64
Guenther et al. 2005, A&A, 442, 1031
Hao, Kouwenhoven & Spurzem, 2013, MNRAS, 433, 867
Hatzes & Cochran, 1992, in *ESO Workshop on High-Res Spectroscopy*, ESO conf. proc. v40, p275
Hatzes et al. 2006, A&A, 457, 335
Hatzes & Zechmeister, 2007, ApJ, 670, L37
Holden et al. 2011, PASP, 123, 14
Keller et al. 2007, PASA, 24, 1
Keller et al. 2014, Nature, 506, 463
Khochfar et al. 2007, ApJ, 668, 115

Kjeldsen & Bedding,1995, A&A, 293, 87
Kriek et al. 2009, ApJ, 700, 221
Le Fèvre et al. 2010, SPIE, 7735, 75
Lovis & Mayor, 2007, A&A, 472, 657
Lu et al. 2009, ApJ, 690, 1463
Luhman, 2013, ApJ, 767, L1
Martins et al. 2007, A&A, 468, 233
Mauerhan et al. 2010, ApJ, 725, 188
McLure et al. 2013, MNRAS, 432, 2696
Mei et al. 2014, ApJ, arXiv: 1403.7524
Meibom et al. 2013, Nature, 499, 55;
Menci et al. 2008, ApJ, 686, 219
Mitchell & Stewart, 2010, ApJ, 722, 1115
Montalto et al. 2007, A&A, 470, 1137
Montalto et al. 2011, A&A, 535, A39
Navarro et al. 2010, SPIE, 7735, 88
Newman et al. 2013, ApJ, 767, 104
Newman et al. 2014, ApJ, arXiv:1310.6754
Norris et al. 2013, ApJ, 762, 28
Oh et al. 2013, ApJ, 767, 117
Pasquini et al. 2012, A&A, 545, A139
Paumard et al. 2006, ApJ, 643, 1011
Pfalzner, 2013, A&A, 549, A82
Preibisch & Zinnecker, 2007, IAUS237, 270
Puech et al. 2008, MNRAS, 390, 1089
Puech et al. 2010, MNRAS, 402, 903
Quinn et al. 2012, ApJ, 756, L33
Reddy & Steidel, 2009, ApJ, 692, 778
Reid, 2006, STScI Symp. Series, astro-ph/0608298
Roederer et al. 2014, AJ, 147, 136
Rousset et al. 2010, SPIE, 7736, 25
Ryan et al. 2007, ApJ, 668, 839
Rudick et al. 2011, ApJ, 732, 48
Santini et al. 2012, A&A, 538, 33
Sato et al. 2007, ApJ, 661, 527
Schőrk et al. 2009 A&A 507, 817
Starkenburg et al. 2010, A&A, 513, 34
Tadaki et al. 2014, ApJ, 788, L23
Tan et al. 2014, A&A, arXiv:1403.7992
Thatte et al. 2010, SPIE, 7735, 85
Thompson, 2013, MNRAS, 431, 63
Thorsett et al. 1999, ApJ, 523, 763
Throop & Bally, 2005, ApJ, 623, L149
Toft et al. 2012, ApJ, 754, 3
van Boekel et al. 2004, Nature, 432, 479
van der Marel et al. 2013, Science, 340, 1199
Villar-Martín et al. 2011, MNRAS, 418, 2032
Weldrake et al. 2005, ApJ, 620, 1043
Weldrake et al. 2008, ApJ, 674, 1117
York et al. 2000, AJ, 120, 1579
Zeimann et al. 2012, ApJ, 756, 115
Zubovas & King, 2012, MNRAS, 426, 2751